\documentclass[12pt]{article}
\usepackage[dvips]{graphicx}
\usepackage{latexsym,cite}
\newcommand{\tppp}{\tau\to\pi\pi\pi\nu_\tau}
\newcommand{\tpppc}{\tau^-\to\pi^+\pi^-\pi^-\nu_\tau}
\newcommand{\tpppn}{\tau^-\to\pi^-\pi^0\pi^0\nu_\tau}

\newcommand{\Frac}[2]{\frac{\displaystyle #1}{\displaystyle #2}}
\newcommand{\lsim}{\stackrel{<}{_\sim}}
\newcommand{\gsim}{\stackrel{>}{_\sim}}
\renewcommand{\thesection}{\arabic{section}}
\def\mapright#1#2{\smash{
     \mathop{-\!\!\!-\!\!\!-\!\!\!\longrightarrow}\limits^{#1}_{#2}}}

\begin{document}

\thispagestyle{empty}
\begin{titlepage}
\begin{center}
\hfill IFIC/03$-$38\\ 
\hfill FTUV/03$-$1215 \\
\vspace*{3.5cm} 
{\Large \bf $\tppp$ decays in the Resonance Effective Theory}
\vspace*{1.7cm} \\
{ \sc D.\ G\'omez Dumm$^a$, A.\ Pich$^b$, J.\ Portol\'es$^b$}
\vspace*{1.2cm} \\
{\em $^a$ IFLP, Depto.\ de F\'{\i}sica, Universidad Nacional de La Plata, \\
C.C. 67, 1900 La Plata, Argentina \\ $\,$\\
$^b$Departament de F\'{\i}sica Te\`orica, IFIC,
CSIC -- Universitat de Val\`encia \\
Apt. Correus 22085, E-46071 Val\`encia, Spain}
\vspace*{1.7cm}
\begin{abstract}
$\tppp$ decays are analysed within the framework of the resonance 
effective theory of QCD. We work out the relevant Lagrangian that
describes the axial--vector current hadronization contributing to 
these processes, in particular the local $a_1$(1260)--$\rho$(770)--Goldstone 
interactions.  The new coupling constants are constrained by 
imposing the asymptotic behaviour of the corresponding spectral function
within QCD. Hence we compare the theoretical framework with the experimental
data, obtaining a good quality fit from the ALEPH spectral function and 
branching ratio. We also get values for the mass and on--shell width of 
the $a_1$(1260) resonance. In this way we are able to provide
the structure functions that have been measured by OPAL and CLEO-II
and we find an excellent agreement. 
\end{abstract}
\end{center}
\vfill
\hspace*{0.5cm} PACS~: 11.30.Rd, 11.40.-q, 13.35.-r, 14.60.Fg  \\
\hspace*{0.5cm} Keywords~: Hadronic tau decays, effective theories, QCD \\ 
\eject
\end{titlepage}

\section{Introduction}

The decays of the $\tau$ lepton offer an excellent laboratory for the
analysis of various topics in particle physics. In particular, $\tau$
decays into hadrons allow to study the properties of the vector
and axial--vector QCD currents, and provide relevant information on the
dynamics of the resonances entering into the processes. The hadronization
of those currents involves the strong interaction at low energies and,
therefore, non--perturbative features of QCD have to be implemented
properly into their evaluation. The Effective Theory framework has a 
long--history of successful achievements in this task.
\par
At very low energies, typically $E \ll M_{\rho}$ (being $M_{\rho}$ 
the mass of the $\rho$(770), the lightest hadron resonance), chiral perturbation
theory ($\chi$PT) \cite{chpt,GL:85} is the corresponding effective theory
of QCD. However the decays $\tppp$, through their full energy spectrum,
happen to be driven by the $\rho$(770) and $a_1$(1260) resonances, 
mainly, in an energy region where they reach their on--shell peaks. 
In consequence $\chi$PT is not 
directly applicable to the study of the whole spectrum but only to 
the very low energy domain \cite{CFU:96}. Until now the standard way
of dealing with these decays has been to use ${\cal O}(p^2)$ $\chi$PT 
to fix the normalization of the amplitudes in the low energy region
and, accordingly, to include the effects of vector and axial--vector
meson resonances by modulating the amplitudes with {\em ad hoc}
Breit--Wigner functions \cite{KS:90,models}. However
we will see that this modelization is not consistent with ${\cal O}(p^4)$
$\chi$PT, a fact that could potentially spoil any outcome provided by
this procedure.
\par
In the last years several experiments have collected good quality 
data on $\tppp$, such as branching ratios and spectra
\cite{ALE,Exp} or structure functions \cite{WXexp}, and their analysis
within a model--independent framework is highly desirable if
one wishes to collect information on the hadronization of the QCD
currents. The Effective Field Theory still provides the proper tool
to proceed in this study. At energies $E \sim M_{\rho}$ the resonance
mesons are active degrees of freedom that cannot be integrated out,
as in $\chi$PT, and they have to be properly included into the relevant
Lagrangian \cite{coleman}. The procedure is ruled by the chiral
symmetry under $SU(3)_{\mathrm{L}} \otimes SU(3)_{\mathrm{R}}$, that 
drives the interaction of Goldstone bosons (the lightest
octet of pseudoscalar mesons), and the $SU(3)_{\mathrm{V}}$ assignments of the
resonance multiplets. Its systematic arrangement has been put forward
in Refs.~\cite{EGPR:89,EGLPR:90} as the Resonance Chiral Theory (R$\chi$T).
We will attach to this framework and we will complete it, when needed, 
in order to fulfill our requirements. 
\par
A complementary tool, thoroughly intertwined with the R$\chi$T, is the
large number of colours ($N_C$) limit of QCD. 
Strong interactions in the resonance region lack an expansion parameter
that, as in $\chi$PT, could provide a perturbative treatment of the
amplitudes. However it has been pointed out \cite{colour1} that the inverse 
of the number of colours of the gauge group $SU(N_C)$ could accomplish 
this task. Indeed large--$N_C$ QCD shows features that resemble, both
qualitatively and quantitatively, the $N_C=3$ case \cite{colour2}. 
Relevant consequences of this approach are that meson dynamics
in the large--$N_C$ limit is described by the tree diagrams of an 
effective local Lagrangian; moreover, at the leading order, one has to
include the contributions of the infinite number of zero--width 
resonances that constitute the spectrum of the theory.
\par  
In this article we study the $\tppp$ processes within the Resonance Effective
Theory of QCD. We will evaluate the form factor associated to the 
$J^P = 1^+$ piece of the axial--vector current that is, by far, the 
dominant contribution to these processes. To proceed we will work out
the effective Lagrangian that endows the dynamics of the decay at the
leading order in the $1/N_C$ expansion, though we will consider  
the lightest vector and axial--vector octets of resonances only. Moreover
we will provide these with a finite width, due to the fact that they 
do really resonate along the full energy spectrum. In order to improve
the construction of the Effective Theory we will constrain the new
unknown couplings in the Lagrangian by demanding the asymptotic behaviour
of the form factor ruled by perturbative QCD. The theoretical
construction will be followed by the phenomenological analysis of
experimental data.
\par
The article is organized as follows~: The resonance effective theory
is studied in Sect.\ 2. In Sect.\ 3 we consider the $\tppp$ processes
and calculate, within our approach, the relevant hadronic matrix
elements. Sect.\ 4 is devoted to discuss the constraints imposed by 
perturbative QCD on the form factors driven by the resonances, which
lead to a set of relations between the coupling constants. 
The analysis of experimental data within our framework is presented
in Sect.\ 5, while in Sect.\ 6 we sketch our conclusions. An Appendix
recalls the basic features of the very--low energy domain of the
$\tpppc$ process within $\chi$PT.

\section{The Resonance Effective Theory of QCD}

The very low--energy strong interaction in the light quark sector
is known to be ruled by the $SU(3)_{\mathrm{L}} \otimes SU(3)_{\mathrm{R}}$
chiral symmetry of massless QCD implemented in $\chi$PT. The leading
${\cal O}(p^2)$ Lagrangian is~:
\begin{equation}
{\cal L}_{\chi}^{(2)}=\frac{F^2}{4}\langle u_{\mu}
u^{\mu} + \chi _+ \rangle \ ,
\label{eq:op2}
\end{equation} 
where
\begin{eqnarray}
u_{\mu} & = & i [ u^{\dagger}(\partial_{\mu}-i r_{\mu})u-
u(\partial_{\mu}-i \ell_{\mu})u^{\dagger} ] \ , \nonumber \\ 
\chi_{\pm} & = & u^{\dagger}\chi u^{\dagger}\pm u\chi^{\dagger} u\ \ 
\ \ , \ \ \ \ 
\chi=2B_0(s+ip) \; \; ,
\end{eqnarray}
and $\langle \ldots \rangle$ is short for a trace in the flavour space.
The Goldstone octet of pseudoscalar fields
\begin{equation}
\Phi (x) \equiv \, \Frac{1}{\sqrt{2}} \, \sum_{a=1}^8 \, \lambda_a \,
\varphi_a \, 
 = \, 
\pmatrix{{1\over\sqrt 2}\pi^0 \, + 
\, {1\over\sqrt 6}\eta_8
 & \pi^+ & K^+ \cr
\pi^- & - {1\over\sqrt 2}\pi^0 \, + \, {1\over\sqrt 6}\eta_8   
 & K^0 \cr K^- & \bar{K}^0 & - {2 \over\sqrt 6}\eta_8 } 
\label{eq:phi_matrix}
\end{equation}
is realized non--linearly into the unitary matrix in the 
flavour space
\begin{equation}
u(\varphi)=\exp \left\{ i\frac{\Phi}{\sqrt{2}\,F} \right\} \; \; \; ,
\end{equation}
that transforms as
\begin{equation}
u(\varphi)  \to  g_R\, u(\varphi)\, h(g,\varphi)^\dagger
                 = h(g,\varphi)\, u(\varphi)\, g_L^\dagger \; \; ,
\end{equation}
with 
$g \equiv (g_L,g_R) \, \in \, SU(3)_{\mathrm{L}} \otimes SU(3)_{\mathrm{R}}$
and $h(g,\varphi)\,\in \, SU(3)_V$. External hermitian
matrix fields $r_{\mu}$, $\ell_{\mu}$, $s$ and $p$ promote the
global $SU(3)_{\mathrm{L}} \otimes SU(3)_{\mathrm{R}}$ symmetry to a local
one. Interactions with electroweak bosons can be accommodated through
the vector $v_{\mu} = (r_{\mu} + \ell_{\mu}) / 2$ and 
axial--vector $a_{\mu} = (r_{\mu} - \ell_{\mu}) / 2$ fields. The scalar
field $s$ incorporates explicit chiral symmetry breaking through the
quark masses $s = {\cal M} \, + \ldots$, with
${\cal M} =  \mathrm{diag}(m_u,m_d,m_s)$ and, finally,
$F \simeq F_{\pi} \simeq 92.4 \, \mbox{MeV}$ is the pion decay constant and 
$B_0 F^2 = - \langle 0 | \bar{\psi}\psi | 0 \rangle_0$ in the chiral limit.
\par
The final hadron system in the $\tppp$ decays spans a wide energy
region $3 m_{\pi}  \lsim  E  \lsim  M_{\tau}$ that is 
populated by resonances. As a consequence an effective theory 
description of the full energy spectrum requires to include the
resonances as active degrees of freedom and, therefore, to consider
R$\chi$T. This recalls the ideas of effective 
Lagrangians from Ref.~\cite{coleman}. The interaction of the 
Goldstone bosons (the lightest octet of pseudoscalar mesons) with the
resonances is driven by chiral symmetry, while the resonances enter as 
$SU(3)_{\mathrm{V}}$ multiplets. In Ref.~\cite{EGPR:89} the simplest
\footnote{These were including one ${\cal O}(p^2)$ chiral tensor in
the interaction operators. It is not possible to construct an interaction
term with lower order in momenta.}
couplings were introduced using the antisymmetric tensor formulation
to describe the vector and axial--vector resonances but keeping linear 
terms in the latter, i.e. allowing for the couplings of the pseudoscalars
to only one resonance. This formalism proved
successful in accounting, upon integration of resonances,
for the bulk of the low--energy coupling 
constants at ${\cal O}(p^4)$ in $\chi$PT with no need of the addition of local
terms \cite{EGLPR:90}. We will attach to this antisymmetric tensor 
formulation and, accordingly, we will not include the 
${\cal O}(p^4)$ $\chi$PT couplings into our effective action in 
order to avoid
double counting of contributions. The success of R$\chi$T in the description 
of many observables has been remarkable, in particular for the pion 
electromagnetic form factor entering $\tau^- \to \pi^- \pi^0 \nu_{\tau}$
\cite{GP:97}.
\par
Therefore in order to introduce the octet of resonance fields we 
define \cite{EGPR:89}
\begin{equation}
R \equiv \frac{1}{\sqrt{2}} \sum_{a=1}^8 \lambda_a R_a\;, \label{defr}
\end{equation}
where $R_a=V_a,A_a$,	stands for vector and axial--vector meson fields,
respectively, that transform as
\begin{eqnarray}
R & \to & h(g,\varphi)\, R \, h(g,\varphi)^\dagger 
\label{eq:rtrans}
\end{eqnarray}
under the chiral group. The flavour structure of the resonances
is analogous to the Goldstone bosons in Eq.~(\ref{eq:phi_matrix}). 
We also introduce the
covariant derivative
\begin{eqnarray}
\nabla_\mu X &	\equiv &	 \partial_{\mu} X	 + [\Gamma_{\mu}, X] \; \; , \\
\Gamma_\mu & = & \Frac{1}{2} \, [ \,
u^\dagger (\partial_\mu - i r_{\mu}) u +
u (\partial_\mu - i \ell_{\mu}) u^\dagger \,] \; \; ,\nonumber
\end{eqnarray}
for any object $X$ that transforms as $R$ in Eq.~(\ref{eq:rtrans}), like
$u_{\mu}$ and $\chi_{\pm}$.
\par
Let us now restrict the analysis to the case of $V(1^{--})$ and
$A(1^{++})$ resonances, in order to describe the couplings of 
$\rho$(770) and $a_1$(1260) states, respectively.
Moreover we include the lightest $SU(3)_{\mathrm{V}}$ 
octet of both vector and axial--vector resonances only. 
Although the leading $1/N_C$ expansion requires to consider an infinite
number of resonances their contribution happens to be suppressed
as their masses become heavier.
\par
The kinetic
terms for the spin 1 resonances in the Lagrangian read, in the
antisymmetric tensor formulation,
\begin{equation}
{\cal L}_{ \rm kin}^R = -\frac{1}{2} \langle
\nabla^\lambda R_{\lambda\mu} \nabla_\nu R^{\nu\mu} - \frac{M_R^2}{2}
R_{\mu\nu} R^{\mu\nu} \rangle \; \; \; , \; \; \;  \; \; R \, = \, V,A \; ,
\end{equation}
being $M_V$, $M_A$ the mass of the octet of vector and axial--vector
resonances, respectively.
The lowest order interaction Lagrangian
includes three coupling constants \cite{EGPR:89},
\begin{eqnarray}
\label{lag1}
{\cal L}_2^{ \rm V} & = &  \frac{F_V}{2\sqrt{2}} \langle V_{\mu\nu}
f_+^{\mu\nu}\rangle + i\,\frac{G_V}{\sqrt{2}} \langle V_{\mu\nu} u^\mu
u^\nu\rangle  \; \, , \nonumber \\
{\cal L}_2^{ \rm A} & = &  \frac{F_A}{2\sqrt{2}} \langle A_{\mu\nu}
f_-^{\mu\nu}\rangle \;,
\end{eqnarray}
where
$f_\pm^{\mu\nu}  =  u F_L^{\mu\nu} u^\dagger \pm u^\dagger F_R^{\mu\nu}
u$ and $F_{R,L}$ are the field strength tensors associated
with the external right and left fields. All coupling parameters
$F_V$, $G_V$ and $F_A$ are real. Notice that the ${\cal O}(p^2)$ interaction
Lagrangian (\ref{lag1}) does not
allow a tree-level coupling of the form $a_1(1260) \pi\pi\pi$.
\par
In the case of $\tppp$, in addition to one--resonance--exchange diagrams one
has to consider the contribution given by the chain 
$a_1(1260) \to\rho(770)\pi\to
\pi\pi\pi$, which is mediated by both vector and axial--vector
resonances. Thus we need to go one step beyond the analysis in
Ref.~\cite{EGPR:89}, including bilinear terms in the resonance fields that
lead to a coupling of the form $a_1\rho\pi$. For the processes under study,
we only need to
consider terms that include a vertex with one pseudoscalar. This 
goal is achieved with one ${\cal O}(p^2)$ chiral tensor. Hence the most 
general interaction
Lagrangian allowed by the symmetry constraints can be written as
\begin{equation}
\label{eq:lag21}
{\cal L}_2^{\rm VAP} \, = \, \sum_{i=1}^{5} \, \lambda_i \, 
{\cal O}^i_{\rm VAP} \; \; ,
\end{equation}
where $\lambda_i$ are new unknown real adimensional couplings, and
the operators ${\cal O}^i_{\rm VAP}$ are given by
\begin{eqnarray}
\label{lag2}
{\cal O}^1_{\rm VAP} &  = & \langle \,  [ \, V^{\mu\nu} \, , \, 
A_{\mu\nu} \, ] \,  \chi_- \, \rangle \; \; , \nonumber \\
{\cal O}^2_{\rm VAP} & = & i\,\langle \, [ \, V^{\mu\nu} \, , \, 
A_{\nu\alpha} \, ] \, h_\mu^{\;\alpha} \, \rangle \; \; , \\
{\cal O}^3_{\rm VAP} & = &  i \,\langle \, [ \, \nabla^\mu V_{\mu\nu} \, , \, 
A^{\nu\alpha}\, ] \, u_\alpha \, \rangle \; \; ,  \nonumber \\
{\cal O}^4_{\rm VAP} & = & i\,\langle \, [ \, \nabla^\alpha V_{\mu\nu} \, , \, 
A_\alpha^{\;\nu} \, ] \,  u^\mu \, \rangle \; \; , \nonumber \\
{\cal O}^5_{\rm VAP} & =  & i \,\langle \, [ \, \nabla^\alpha V_{\mu\nu} \, , \, 
A^{\mu\nu} \, ] \, u_\alpha \, \rangle \nonumber \; \; ,
\end{eqnarray}
with $h_{\mu \nu} = \nabla_{\mu} u_{\nu} + \nabla_{\nu} u_{\mu}$.
We emphasize that this set is a complete basis for 
constructing vertices with only one
pseudoscalar; for a larger number
of pseudoscalars additional operators may emerge.
As we are only interested in tree level diagrams, the equation of motion
arising from ${\cal O}(p^2)$ $\chi$PT, 
\begin{equation}
h_\alpha^{\;\alpha} \, = \,  i \, \chi_-  \, - \, \Frac{i}{3} 
\langle \chi_{-} \rangle \; ,
\end{equation}
has been used in  
${\cal L}_2^{\rm VAP}$ in order to eliminate one of the 
possible operators.
\par
In summary we will proceed in the following by considering the
relevant Resonance Chiral Theory given by~:
\begin{equation}
\label{eq:ret}
{\cal L}_{\rm R\chi T} \, = \, 
{\cal L}_{\chi}^{(2)} \, + \, {\cal L}_{\rm kin}^{\rm V} \, + \, 
{\cal L}_{\rm kin}^{\rm A}
\, + \, {\cal L}_2^{\rm V} \, + \, {\cal L}_2^{ \rm A} \, +
\, {\cal L}_2^{\rm VAP}
\; \; .
\end{equation}
It is important to notice that 
${\cal L}_{\rm R\chi T}$ is not an effective theory of QCD
for arbitrary couplings in the interaction terms 
${\cal L}_2^{\rm V/A}$ and ${\cal L}_2^{\rm VAP}$.
These carry some symmetry properties of QCD, implemented in the 
operators, that do not give any information on the coupling
constants. However we will see in Section 4 that well accepted
dynamical properties of the underlying theory provide some information
on those couplings. This is as far as we can go without
including modelizations.

\section{Axial--vector current form factors in $\tppp$}

In this Section we describe the $\tppp$ decays in the isospin limit and proceed
to calculate the relevant hadronic matrix element within the resonance
effective theory given by ${\cal L}_{\rm R\chi T}$. 
The decay amplitude for the $\tpppc$ and $\tpppn$ processes 
can be written as
\begin{equation} {\cal M}_{\pm} \; = \; - \,
\frac{G_F}{\sqrt{2}} \, V_{ud} \, \bar u_{\nu_\tau}
\gamma^\mu\,(1-\gamma_5) u_\tau\, T_{\pm \mu} \; \; ,
\end{equation}
where $V_{ud}$ is an element of the
Kobayashi--Maskawa matrix  and $T_{\pm \mu}$ is the hadronic matrix
element of the participating $V_{\mu} - A_{\mu}$ QCD currents. In the isospin
limit there is no contribution of the vector current to these processes and,
therefore, only the axial--vector current $A_\mu$ appears~:
\begin{equation}
T_{\pm \mu}(p_1,p_2,p_3) \; = \; 
 \langle \, \pi_1(p_1)\pi_2(p_2)\pi^{\pm}(p_3) \, | \,  A_\mu(0) \, | \, 
 0 \, \rangle \; \; ,
\label{matelem}
\end{equation}
being $\pi^+$ the one in $\tpppc$ and  $\pi^-$ that in $\tpppn$ (in the 
following we will omit
the subscript $\pm$ when speaking generally). The hadronic
tensor can be written in terms of three form factors, $F_1$, $F_2$ and
$F_P$, as \cite{KM:92}
\begin{equation}
T^{\mu} \; = \; V_1^\mu\,F_1 \, + \,  V_2^\mu\,F_2 \, + \,  
Q^\mu\,F_P \; \; ,
\label{tmu}
\end{equation}
where
\begin{eqnarray}
V_1^\mu & = & \left( \, g^{\mu \nu} \, - \, 
\Frac{Q^{\mu} Q^{\nu}}{Q^2} \, \right) \, ( \, p_1 - p_3 \, )_{\nu} \; \; ,
\nonumber \\
V_2^\mu & = & \left( \, g^{\mu \nu} \, - \, 
\Frac{Q^{\mu} Q^{\nu}}{Q^2} \, \right) \, ( \, p_2 - p_3 \, )_{\nu} \; \; ,
\nonumber \\
Q^\mu & = & \,  p_1^\mu + p_2^\mu + p_3^\mu\; \; .
\end{eqnarray}
\par
Now if $\Gamma$ stands for the $\tppp$ decay width, the spectral function
$d\Gamma/dQ^2$ can be written as
\begin{equation}
\frac{d \, \Gamma}{dQ^2} = \frac{G_F^2\, |V_{ud}|^2}
{128\, (2\pi)^5\, M_\tau}\;
\left( \, \Frac{M_\tau^2}{Q^2}-1 \, \right)^2 \int ds\, dt\,
\left[ W_{SA} + \frac{1}{3}\left(1 \, + \, 2 \,\frac{Q^2}{M_\tau^2}\right)
W_A \right] \;,
\label{dgdq}
\end{equation}
where $s=(p_1+p_3)^2$, $t=(p_2+p_3)^2$, and the hadronic structure
functions $W_{SA}$ and $W_A$ are given by
\begin{eqnarray}
W_{SA} & = & (Q^\mu\,F_P)\,(Q_{\mu}\,F_P^{\ast}) \;\; = \;\;
Q^2\, |F_P|^2 \; \; , \nonumber \\
W_A & = & -\, (V_1^\mu\,F_1 \, + \,  V_2^\mu\,F_2)
\,(V_{1\mu}\,F_1 \,  + \, V_{2\mu}\,F_2)^\ast \; \; .
\label{wa}
\end{eqnarray} 
The phase--space
integrals extend over the region allowed for a three--pion state with a
center--of--mass energy $\sqrt{Q^2}$:
\begin{equation}
\int ds\, dt\, \equiv
\int_{4m_\pi^2}^{(\sqrt{Q^2} - m_\pi)^2} ds\;\;
\int_{t_{-}(s)}^{t_+(s)} dt\;\;,
\end{equation}
where
\begin{equation}
t_\pm = \frac{1}{4\,s}\left\{
(Q^2-m_\pi^2)^2 - \left[\lambda^{1/2}(Q^2,s,m_\pi^2)
\mp \lambda^{1/2}(s,m_\pi^2,m_\pi^2)\right]^2\right\}\; \; ,
\end{equation}
with $\lambda(a,b,c)=(a+b-c)^2-4ab$. We have neglected here the mass of the
$\nu_{\tau}$.
\par
In the decomposition of $T_\mu$ in Eq.~(\ref{tmu}), the form factors $F_1$ and 
$F_2$ have a transverse structure in the 
total hadron momenta $Q_{\mu}$ and drive a
$J^P=1^+$ transition. Bose symmetry under interchange of the
two identical pions in the final state demands that 
$F_1(Q^2,s,t) = F_2(Q^2,t,s)$. Meanwhile $F_P$ accounts for a $J^P=0^-$
transition that carries pseudoscalar degrees of 
freedom. As a consequence, the conservation of the axial--vector current 
in the chiral limit imposes that $F_P|_{m_{\pi}=0} = 0$, and the scalar 
form factor must vanish with the square of the pion mass. Hence its
contribution to the decay processes will be very much suppressed. 

\subsection{Evaluation of the matrix amplitude}

We proceed now to calculate the hadronic amplitudes $T_{\pm \mu}$ as given
by our Resonance Effective Theory described in Section 2. Within the 
large--$N_C$ framework, one should evaluate all tree--level diagrams
generated by ${\cal L}_{\rm R\chi T}$ that contribute to
the decays \footnote{We remind that 
though large--$N_C$ enforces the 
contributions of the infinite spectrum of QCD resonances, we only include
the lightest octet of vector and axial--vector mesons. The phenomenology
seems to support well this approach.}.
\par
\begin{figure}[t]
\vspace*{0.5cm}
\centerline{
   \includegraphics[height=3.2truecm]{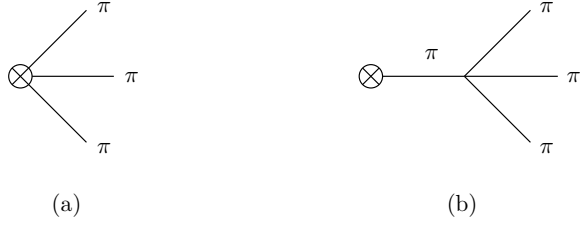}
   } \caption{Diagrams contributing to the hadronic amplitude $T_{\pm \mu}$
    at ${\cal O}(p^2)$ in $\chi$PT. The crossed circle indicates 
    the insertion of the
    axial--vector current.}
\label{fig:primera}
\end{figure}

In the low $Q^2$ region, the matrix element in Eq.\ (\ref{matelem}) can be
calculated using $\chi$PT. At ${\cal O}(p^2)$ one has two contributions, 
arising from the diagrams (a) and (b) of Fig.\ \ref{fig:primera}. 
The sum of both graphs yields
\begin{eqnarray}
T_{\pm\mu}^{\chi} & = & \frac{2\sqrt{2}}{3 F}\left[\pm
\left(g_{\mu\nu}-\frac{Q_\mu Q_\nu}{Q^2-m_\pi^2}\right)\, (2\,
p_3-p_2-p_1)^\nu - \frac{\kappa_\pm\,m_\pi^2}{Q^2-m_\pi^2}
\;Q_\mu \right] \nonumber \\
& = & \mp \frac{2\sqrt{2}}{3 F}\left\{ V_{1\mu} + V_{2\mu} -
\frac{m_\pi^2\, [3\,(u-m_\pi^2)- Q^2\,(1\pm
2\kappa_\pm)]}{2Q^2(Q^2-m_\pi^2)}\;\,Q_\mu \right\} \, \; , 
\label{lowq}
\end{eqnarray}
where we have defined $u=(p_1+p_2)^2=Q^2-s-t+3m_\pi^2$, $\kappa_+ = 1$,
$\kappa_-=1/2$. As expected from PCAC, it can be seen that the amplitude
is transverse in the limit where the pion mass is neglected.

\par
In the limit of low $Q^2$ (i.e. $Q^2 \ll M_R^2$) the resonance fields 
can be integrated
out, and their contributions should reduce to the corrections to
Eq.~(\ref{lowq}) obtained from the effective ${\cal O}(p^4)$ interactions
and higher order
couplings in the standard $\chi$PT Lagrangian. In hadronic $\tau$ decays,
however, we need to implement the resonance degrees of freedom explicitly
because $Q^2 \sim M_R^2$.
As stated in Section 2 we assume here that the resonances 
$\rho$(770) and $a_1$(1260) give the dominant contributions to $\tppp$ 
decays. Consequently, in addition to the usual
$\chi$PT diagrams leading to Eq.\ (\ref{lowq}) we include 
resonance--mediated contributions
to the amplitude, to be evaluated through the interacting terms
${\cal L}_2^{\rm V/A}$ and ${\cal L}_2^{\rm VAP}$ of the resonance effective
theory. 
\par
\begin{figure}[t]
\centerline{
   \includegraphics[height=3.8truecm]{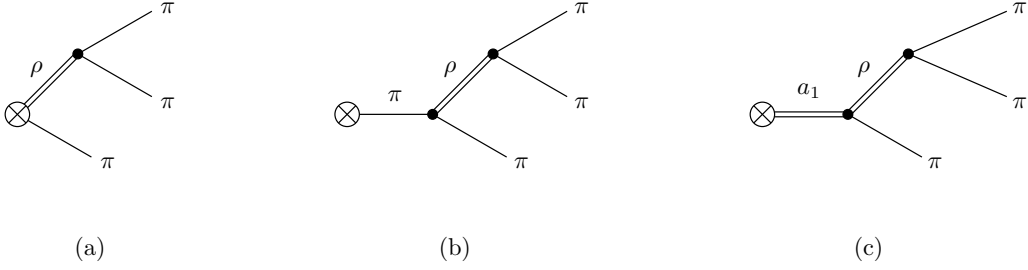}
   } \caption{Resonance--mediated diagrams contributing to the hadronic
amplitude $T_{\pm \mu}^{(1R)}$, (a) and (b), and $T_{\pm \mu}^{(2R)}$, 
(c).}
\label{fig:segunda}
\end{figure}

The relevant diagrams to be taken into account are those shown in 
Fig.\ \ref{fig:segunda}.
Let us start by considering the contributions given by the graphs in 
Figs.\ \ref{fig:segunda} (a) and (b), which involve only one 
intermediate resonance. From the
interaction Lagrangian in Eq.\ (\ref{lag1}), the sum of both graphs leads
to
\begin{equation}
\label{eq:t1r}
T_{\pm\mu}^{(1R)} = \mp \frac{\sqrt{2}\,F_V\,G_V}{3\,F^3}
\,\left\{\,\alpha_1(Q^2,s,t)\,V_{1\mu} + \alpha_1(Q^2,t,s)\,V_{2\mu}
+ [\alpha_2(Q^2,s,t)+\alpha_2(Q^2,t,s)]\,Q_\mu\right\} \, ,
\end{equation}
where we have defined
\begin{eqnarray}
\alpha_1(Q^2,s,t) & \equiv & - \, 3 \,  \Frac{s}{s-M_V^2} \, + \, 
\left( \Frac{2 G_V}{F_V} - 1 \right) \, \left\lbrace 
\, \Frac{2 Q^2-2s-u}{s-M_V^2} \, + \, \Frac{u-s}{t-M_V^2} \, \right\rbrace 
\;\;, \nonumber \\
\alpha_2(Q^2,s,t) & \equiv & 3 \, \Frac{G_V}{F_V} \, \Frac{s}{Q^2} \, 
\Frac{m_{\pi}^2}{Q^2-m_{\pi}^2} \, \Frac{u-t}{s-M_V^2} \; \; .
\label{ff1r}
\end{eqnarray}
As expected, the pseudoscalar form factor $\alpha_2(Q^2,s,t)$ is found to 
vanish in the chiral limit.
\par

The two--resonance contribution in Fig.\ \ref{fig:segunda} (c) 
can be evaluated taking into account the chiral couplings 
from ${\cal L}_2^{\rm VAP}$ in Eq.\ (\ref{lag2}). We obtain~:
\begin{equation}
\label{eq:t2r}
T^{(2R)}_{\pm\mu} = \pm\frac{4\, F_A G_V}{3\,F^3} \, \Frac{Q^2}{Q^2-M_A^2}
\,\left[\,
\beta(Q^2,s,t)\,V_{1\mu} + \beta(Q^2,t,s)\,V_{2\mu}\right] \; \; ,
\end{equation}
where
\begin{equation}
\beta(Q^2,s,t) \, \equiv \, 
- \, 3 \, (\lambda' + \lambda'') \, \Frac{s}{s-M_V^2} \, 
+ \, \, F(Q^2,s) \, \Frac{2 Q^2 + s - u}{s-M_V^2} \, 
+ \,  F(Q^2,t) \, \Frac{u-s}{t-M_V^2} \, \; \; ,
\label{ff2r}
\end{equation}
with
\begin{eqnarray} \label{eq:lambda0}
F(Q^2,s) & = & - \,\lambda_0\, \Frac{m_\pi^2}{Q^2} \, +  \, 
\lambda'\, \Frac{s}{Q^2} \, + \,  \lambda'' \; \; , 
\nonumber \\
\lambda_0 & = & -\,\frac{1}{\sqrt{2}} \left[ 4\,\lambda_1 \, +
\lambda_2 + \frac{\lambda_4}{2} + \lambda_5 \right] \; \; , \nonumber \\
\lambda' & = & \frac{1}{\sqrt{2}}\, (\lambda_2-\lambda_3 + \frac{\lambda_4}{2}
+ \lambda_5 ) \; \; , \nonumber \\
\lambda'' & = & \frac{1}{\sqrt{2}}\, (\lambda_2 - \frac{\lambda_4}{2} -
\lambda_5 ) \;.
\end{eqnarray}
Notice that the form factor $\beta(Q^2,s,t)$ is written in terms of only
three combinations of the unknown couplings in 
${\cal L}_2^{\rm VAP}$, namely $\lambda_0$, $\lambda'$ and
$\lambda''$.
\par
The final result from our evaluation gives, accordingly, the addition
of all the amplitudes considered here~:
\begin{equation}
T_{\pm \mu} \, = \, T_{\pm \mu}^{\chi} \, + \, T_{\pm \mu}^{(1R)} \,
+ \, T_{\pm \mu}^{(2R)}  \; \; ,
\end{equation}
and, for later use, if we only consider the $J^P=1^+$ mediated
form factors $F_1$ and $F_2$ we will specify $T_{\pm \mu} \, |_{1^+}$.

\subsection{Implementation of off--shell widths}

The form factors in Eqs.\ (\ref{ff1r}) and (\ref{ff2r}) include 
zero--width $\rho$(770) and
$a_1$(1260) propagator poles, leading to divergent phase--space integrals 
in the calculation of the $\tppp$ decay width as the kinematical variables
go along the full energy spectrum. The result can be regularized
through the inclusion of resonance widths, which means to go beyond the
leading order in the $1/N_C$ expansion, and implies the introduction of
some additional theoretical inputs. Moreover both contributing resonances,
especially the $a_1$(1260), are rather wide. Hence energy--dependent
widths should be included in order to handle their off--shell character.  
This issue has been analysed in detail within the resonance
chiral effective theory in Ref.~\cite{GDPP:00}. For the case of spin-1 vector
resonances, it is seen that one can define the off--shell width by taking
into account the pole of the two--point function of the corresponding
vector current that arises from the resummation of those diagrams that
include an absorptive contribution of two pseudoscalars in the $s$ channel. 
The width is defined then as the imaginary part of this pole. 
In the case of the 
$\rho$(770) meson this procedure leads to the result
\begin{equation}
M_V \, \Gamma_\rho(s) \, = \,  \Frac{M_V^2 \,s}{96\,\pi\,F^2}
\left[\sigma_\pi^3\,\theta(s-4\,m_\pi^2) + \frac{1}{2}\,
\sigma_K^3\,\theta(s-4\,m_K^2)\right]\;, \label{rhow}
\end{equation}
where $\sigma_P=\sqrt{1-4m_P^2/s}$, and $\theta(x)$ is the step function.
In the case of the $a_1$(1260) meson the construction following the 
definition above, though well defined, is much more involved. 
It would amount to evaluate the axial--vector--axial--vector current
correlator with absorptive cuts of three pions within the resonance
effective theory. This corresponds to a non--trivial two--loop calculation
within a theory whose regularization is still not well defined. This
calculation is far beyond our scope. In order to perform the phenomenological 
analysis we will introduce a chiral--based off--shell width for the
$a_1$(1260) resonance that endows the appropriate kinematical and
dynamical features~:
\begin{equation}
M_A \, \Gamma_{a_1}(Q^2) = M_A \, \Gamma_{a_1}(M_{A}^2)\,
\Frac{\phi(Q^2)}{\phi(M_{A}^2)} \, \left(
\Frac{M_{A}^2}{Q^2}\right)^\alpha \, \theta(Q^2-9\,m_\pi^2)\;,
\label{a1w}
\end{equation}
where 
~:
\begin{equation}
\phi(Q^2) \, = \, Q^2 \int ds\,dt\;
\bigg\{V_1^2\,|BW_\rho(s)|^2 + V_2^2\,|BW_\rho(t)|^2 +
 2\,(V_1\cdot V_2)\,{\rm Re}[BW_\rho(s)\,BW_\rho(t)^\ast]\;
 \bigg\}\; \; ,
\end{equation}
and
\begin{equation}
BW_\rho(s) = \frac{M_V^2}{M_V^2-s-i\,M_V \, \Gamma_\rho(s)}  
\end{equation}
is the usual Breit--Wigner function for the $\rho$(770) meson resonance shape,
and the energy--dependent width $\Gamma_\rho(s)$ is given by
Eq.~(\ref{rhow}).
\par
In conclusion, while the off-shell width of the $\rho$(770) meson does not
introduce any additional free parameters on our result for 
$T_{\pm \mu}$, the $a_1$(1260) width includes two new ones, namely the
on--shell width
$\Gamma_{a_1}(M_A^2)$ and the parameter $\alpha$ ruling the $Q^2$ behaviour
in Eq.~(\ref{a1w}).

\section{Asymptotic behaviour and QCD constraints}

As commented at the end of Section 2, our Lagrangian
${\cal L}_{\rm R\chi T}$ in Eq.~(\ref{eq:ret}) does not
provide an effective theory of QCD for arbitrary values of its
couplings. The only tool used in its construction were relevant symmetry
properties of QCD, but it is clear that the full theory should
unambiguously predict the values of the coupling constants.
Although it is not known how to achieve this goal from first principles, 
several ideas based on matching procedures have been developed
\cite{EGLPR:90,RF03}. In this Section we show how to proceed 
in order to obtain information on the, a priori, unknown couplings
in ${\cal L}_2^{\rm VAP}$.
\par
The QCD ruled short--distance behaviour of the vector form 
factor in the large--$N_C$ limit (approximated with only one octet
of vector resonances) constrains the couplings of 
${\cal L}_2^{\rm V}$ in Eq.~(\ref{lag1}), which must
satisfy \cite{EGLPR:90}~:
\begin{eqnarray}
\label{fvgv}
1 \, - \, \Frac{F_V \, G_V}{F^2} & = & 0 \; \; . 
\end{eqnarray} 
In addition, the first Weinberg sum rule \cite{W:67}, in the limit 
where only the lowest narrow resonances contribute to the vector 
and axial--vector spectral functions, leads to
\begin{equation}
\label{fvfa}
F_V^2 - F_A^2 = F^2 \;.
\end{equation}
In Ref.~\cite{EGLPR:90} it was also noticed that there is an additional
constraint coming from the axial form factor, namely~:
\begin{equation} \label{eq:dubio}
2 F_V G_V \, - \, F_V^2 \,  = 0 \;  ,
\end{equation}
that is very well satisfied by the phenomenology. However this relation
is not necessarily true when we consider the inclusion of 
vector--pseudoscalar--axial
couplings as given by ${\cal L}_2^{\rm VAP}$. In the following we are going
to enforce the relation given in Eq.~(\ref{eq:dubio}) and we will see
what happens when we relax this last condition. From 
Eqs.~(\ref{fvgv},\ref{fvfa},\ref{eq:dubio}), the three
couplings $F_V$, $G_V$ and $F_A$ in Eq.~(\ref{lag1})
can be written in terms of the pion decay constant~: 
$F_V = \sqrt{2} F$, $G_V = F/\sqrt{2}$ and $F_A = F$. 
We will adopt these results throughout this paper, unless stated otherwise.
\par
An analogous analysis should be done in the case of
the axial two--point function $\Pi_A^{\mu\nu}(Q^2)$, which plays
in $\tppp$ processes the same role than the vector--vector current 
correlator does in the $\tau \rightarrow \pi \pi \nu_{\tau}$ decays, driven
by the vector form factor. The goal will be to obtain QCD--ruled 
constraints on the new couplings of 
${\cal L}_2^{\rm VAP}$, Eq.~(\ref{eq:lag21}), similar to those 
obtained above. As these couplings
do not depend on the Goldstone masses we will work in the chiral limit but
our results will apply for non--zero Goldstone masses too.
In the chiral limit the $\Pi_A^{\mu\nu}(Q^2)$
correlator becomes transverse, hence we can write
\begin{equation}
\Pi_A^{\mu\nu}(Q^2)=(Q^\mu Q^\nu-g^{\mu\nu}Q^2)\,\Pi_A(Q^2)\,.
\end{equation}
As in the case of the pion and axial form factors, the function
$\Pi_A(Q^2)$ is expected to satisfy an unsubtracted dispersion relation.
This implies a constraint for the $J=1$ spectral function Im$\Pi_A(Q^2)$
in the asymptotic region, namely
\begin{equation}
{\rm Im}\Pi_A (Q^2)\,\,
\mapright{\; \; \;  Q^2 \rightarrow \infty \; \; \; }{} \; \; 
0 \; \; .
\label{constr}
\end{equation}
Now, taking into account that each intermediate state carrying the appropriate
quantum numbers yields a positive contribution to Im$\Pi_A (Q^2)$, we have
\begin{equation}
{\rm Im}\Pi_A (Q^2)\,\geq\,
-\frac{1}{3\,Q^2} \int d\Phi\; \left(T^\mu|_{1^+}\right)
\, \left(T_\mu|_{1^+}\right)^\ast\,,
\end{equation}
$d\Phi$ being the differential phase space for the three--pion state. The
constraint in Eq.\ (\ref{constr}) then implies
\begin{equation}
\lim_{Q^2\to\infty}\;
\int_0^{Q^2} ds\; \int_0^{Q^2-s} dt\; \frac{W_A}{(Q^2)^2} = 0 \,,
\label{cond}
\end{equation}
where $W_A$ is the structure function defined in Eq.~(\ref{wa}). 
It can be seen that the condition in Eq.~(\ref{cond}) is not satisfied 
in general for arbitrary values of the
coupling constants in the chiral interaction Lagrangian. Indeed, once
$F_V$, $F_A$ and $G_V$ have been fixed by Eqs.~(\ref{fvgv}), (\ref{fvfa})
and (\ref{eq:dubio}),
we find that the constants $\lambda'$ and $\lambda''$ appearing in the
two--resonance contribution to the hadronic tensor are required to satisfy
the relations
\begin{eqnarray}
2 \, \lambda' \, - \, 1 & = & 0 \; \;, \nonumber \\
\lambda'' & = & 0 \; \;.
\label{rellam}
\end{eqnarray}
Now we can come back to our results for $T_{\pm \mu}$ in Section 3
and enforce the QCD driven constraints on the couplings of our Effective
Theory as given by Eqs.~(\ref{fvgv},\ref{fvfa},\ref{eq:dubio},\ref{rellam}). 
We have~:
\begin{equation}
\label{3/2}
\left.\left. T_{\pm\mu}^{\chi}\right|_{1^+} \, + \, 
 T_{\pm\mu}^{(1R)}\right|_{1^+} \, = \, 
\mp \,\frac{2\sqrt{2}}{3 F} \left[ \left( \, 1 \, - \, \Frac{3}{2} \, 
\Frac{s}{s-M_V^2} \,  \right) \,  V_{1\mu} \,  + \, 
 \left( \,  1 \, - \, \Frac{3}{2} \,  \Frac{t}{t-M_V^2} \,  \right) \, 
 V_{2\mu} \, \right]
\end{equation}
and
\begin{equation}
T_{\pm\mu}^{(2R)} \, = \,  \pm\,\Frac{2\sqrt{2}}{3\,F} \,
\Frac{Q^2}{Q^2-M_A^2} \, 
\,\left[\,\beta(Q^2,s,t)\,V_{1\mu} + \beta(Q^2,t,s)\,V_{2\mu}\right]\;,
\end{equation}
where
\begin{eqnarray}
\beta(Q^2,s,t) \, & \equiv & \, 
- \, \Frac{3}{2} \,  \Frac{s}{s-M_V^2} \, + F(Q^2,s) \, 
\Frac{2 Q^2 + s - u}{s-M_V^2} \, + \, F(Q^2,t)  \, 
\Frac{u-s}{t-M_V^2} \; \; , \nonumber \\
F(Q^2,s) \, & = & \, \left( \, \Frac{s}{2 Q^2} \, 
- \,  \lambda_0 \, \Frac{m_{\pi}^2}{Q^2} \, \right) \; \; .
\end{eqnarray}
Thus, it turns out that the hadronic amplitude can be written in terms of
only one unknown coupling parameter, namely $\lambda_0$.
\par
As mentioned in Section 2, the resonance exchange approximately saturates 
the phenomenological values of the ${\cal O}(p^4)$ couplings in the standard
$\chi$PT Lagrangian. This allows to relate both schemes in the low energy
region, and provides a check of our results in the limit $Q^2\ll
M_V^2$. We have performed this check (see Appendix A), verifying the 
agreement between our expression Eq.\ (\ref{3/2}) and the result 
obtained within $\chi$PT in Refs.\ \cite{CFU:96,DFM:94} coming from
saturation by vector meson resonances of the ${\cal O}(p^4)$ couplings~:
\begin{equation}
\label{chpt}
\left. T_{\pm\mu}^{\chi PT}\right|_{1^+}\, = \,
\mp \,\frac{2\sqrt{2}}{3 F}
\left[ \left( 1 + \frac{3\,s}{2\,M_V^2} \right) V_{1\mu} +
\left( 1 + \frac{3\,t}{2\,M_V^2} \right) V_{2\mu} \right]\,
+ \mbox{ chiral loops }  \, + \, {\cal O}(p^6) \; \;.
\end{equation}
\par
As an aside, it is worth to point out that this low--energy behaviour is not
fulfilled by all phenomenological models proposed in the literature. In
particular, in the widely used model by K\"uhn and Santamaria (KS)
\cite{KS:90} the hadronic amplitude satisfies
\begin{equation}
T_{\pm\mu}^{(KS)}\;\;
\mapright{\; \; s,t\,\ll\, M_V^2 \; \; }
\, \mp \,\Frac{2\sqrt{2}}{3 F}
\left[ \left( 1 + \frac{s}{M_V^2} \right) V_{1\mu} +
\left( 1 + \frac{t}{M_V^2} \right) V_{2\mu} \right] \;.
\label{ks}
\end{equation}
Thus, while the lowest order behaviour is correct (it was constructed
to be so), it is seen that the KS model fails to reproduce the 
$\chi$PT result at the next--to--leading order. Accordingly this model
is not consistent with the chiral symmetry of QCD.

\section{Phenomenology of $\tppp$ processes}

In this Section we discuss the ability of our Effective Theory to
describe the experimental observations in $\tppp$ decays, taking into
account the data obtained from the measurements of $\tppp$ spectra,
branching ratios and structure functions.
\par
Before proceeding let us specify which are the parameters left
unknown in our evaluation of the hadronic matrix amplitude. To analyse
the experimental data we will only consider the dominating $J^P=1^+$ driven 
axial--vector form factors, being the pseudoscalars suppressed by pion 
mass factors. Hence we notice that 
$T_{+ \mu} |_{1^+}\, = \, - \, T_{- \mu} |_{1^+}$ and,
accordingly, we have the same predictions for both 
$\tpppc$ and $\tpppn$ processes in the isospin limit.
As we have shown in the previous Section, the requirement of the 
proper asymptotic behaviour of vector and axial--vector spectral 
functions consistent with QCD imposes several constraints on the 
coupling constants of our Lagrangian ${\cal L}_{\rm R\chi T}$ 
in Eq.~(\ref{eq:ret}) and, as a consequence, the hadronic amplitude 
for the decays $\tppp$ only involves one adimensional unknown combination 
of coupling
constants, $\lambda_0$. In addition, we have three remaining parameters,
related with
the $a_1$(1260) resonance~: these are its mass $M_A$, its on--shell
width $\Gamma_{a_1}(M_A^2)$ and the exponent $\alpha$, which has been
introduced in 
Eq.\ (\ref{a1w}) to account for the dominating energy dependence of the 
off-shell $a_1$(1260) width. The mass of the vector octet $M_V$, 
to be identified with the $\rho$(770) mass, is 
better obtained from the vector form factor of the pion 
\cite{GP:97,PP01} and we take here $M_V = 775.1 \,\mbox{MeV}$. 
In this way, we deal with four free parameters to fit the experimental data.

\subsection{Fit to the ALEPH branching ratio and spectral function}

In order to check the consistency of the available experimental results
with the theoretical description provided by our chirally-based resonance
framework we have carried out a fit of the unknown parameters entering
the amplitudes.
We have chosen to fit the experimental values for the $\tpppc$ branching
ratio and normalized spectral function obtained by ALEPH~\cite{ALE},
which quotes the unfolded spectrum for the decay including the
corresponding covariance matrix. The data are collected in 57
equally spaced bins with a
$3\pi$ squared invariant mass ranging from $Q^2=0.275$ GeV$^2$ to $Q^2=3.075$
GeV$^2$. We have noticed that bins $n=2$ and $n=54$ produce anomalously
high $\chi^2$ contributions, due to their tiny errors, and we have 
discarded them. In addition we have fitted the total branching ratio 
to the value quoted by ALEPH \cite{ALE}, $\mbox{BR}(\tpppc)=(9.15\pm 0.15)$\%, 
which has been introduced as an additional fitting point. In order to 
control possible fake results the
procedure has been carried out both with the MINUIT package \cite{MINUIT} 
and with an independent minimization procedure. 
As output of our fit we find the four parameter set
\begin{eqnarray}
\label{fit}
\lambda_0 & = & 11.9 \pm 0.4 \; \; , \nonumber \\
\alpha & = & 2.45 \pm 0.15 \; \; , \nonumber \\
M_{A} & = & (1.204 \pm 0.007) \rm{\ GeV} \; \; , \nonumber \\
\Gamma_{a_1}(M_A^2) & = & (0.48 \pm 0.02) \rm{\ GeV} \; \;,
\end{eqnarray}
\begin{figure}[t]
\vspace*{0.6cm}
\centerline{
   \includegraphics[width=11truecm,angle=-90]{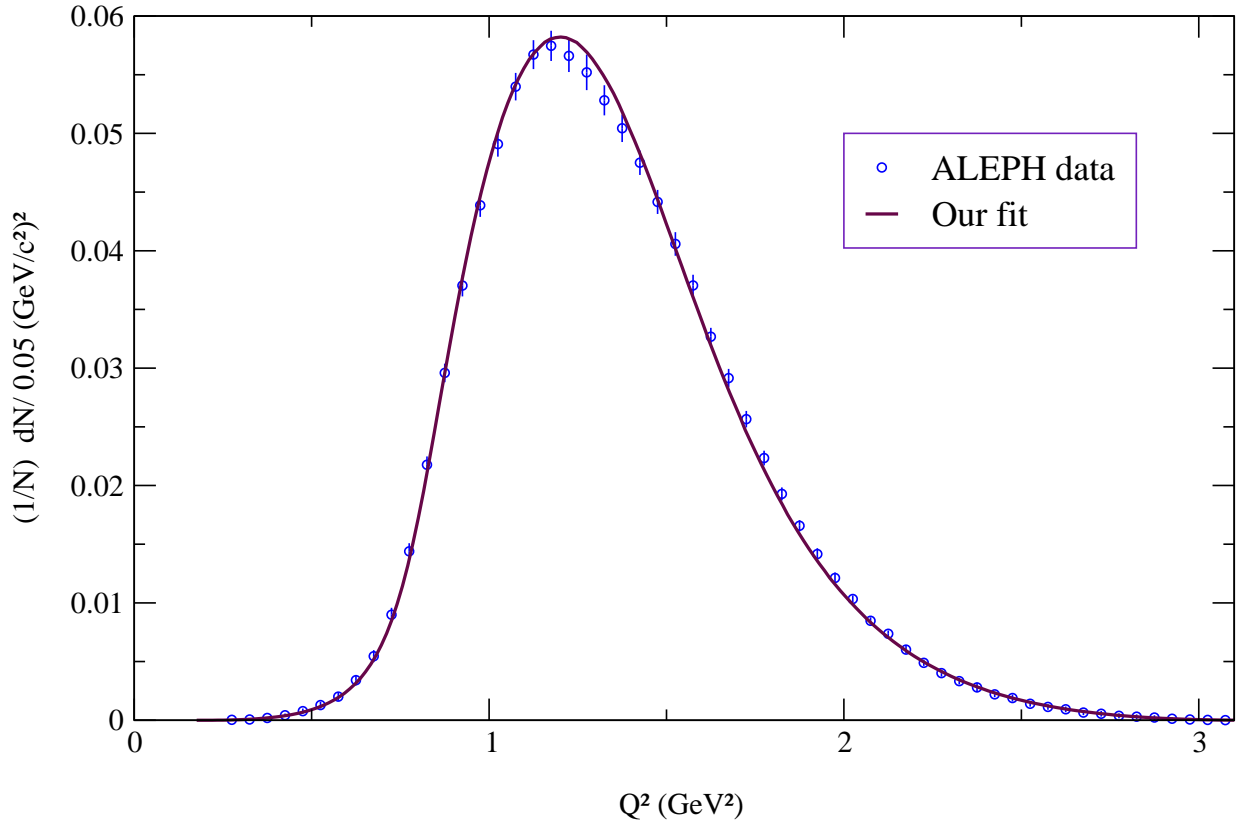}
   } \caption{Fit to the ALEPH data for the normalized $\tpppc$ spectral
function.}
\label{fig:tercera}
\vspace*{0.3cm}
\end{figure}
\hspace*{-0.3cm}
with $\chi^2/\rm{d.o.f.} = 64.5/52$ (these numbers include only statistical
errors). We find that the values of the parameters in Eq.\ (\ref{fit}) 
lead to $\mbox{BR}(\tpppc)=(9.1 \pm 0.3)$\%. The theoretical
spectrum arising from the central values of our fit, together with ALEPH
data, are represented in Fig.~\ref{fig:tercera}. It can be seen that the 
agreement is good, if one considers the very small errors in the
experimental data. On the other hand, the values obtained for the
$a_1$(1260) mass and width are in good agreement with the world average 
values quoted by the Particle Data Group \cite{PDG}.
\par
Taking into account the value of $\alpha$ in Eq.\ (\ref{fit}), we
have carried out three additional three-parameter fits in which $\alpha$ has
been fixed to the values 2, 2.5 and 3, respectively. The
results are shown in Table I. Although the results clearly prefer a value
of $\alpha = 2.5$, one can see that the quality of the fits is
reasonably similar to that of the four-parameter case, 
and the values of $\lambda_0$, $M_{A}$ and $\Gamma_{a_1}(M_A^2)$ are not 
significantly modified.
Thus it can be concluded that a more sophisticated evaluation of the
behaviour of the off-shell $a_1$(1260) width should not imply a major 
qualitative and quantitative changes in our global analysis.

\begin{table}[htb]
\begin{center}
  \begin{tabular}{l||c|c|c|} 
    \cline{2-4}
 & $\alpha = 2$ & $\alpha = 2.5$  & $\alpha = 3$ \\
    \hline
$\lambda_0$ & $12.7 \pm 0.4$ & $11.9 \pm 0.3$ & $11.9 \pm 0.3$ \\
$M_{A}$ & $1.242 \pm 0.004$ & $1.203 \pm 0.003$ & $1.170 \pm 0.002$ \\
$\Gamma_{a_1}(M_{A}^2)$ & $0.55 \pm 0.02$ & $0.48 \pm 0.02$ 
& $0.44 \pm 0.01$ \\
$\chi^2/\rm{d.o.f.}$ & 83.1/53 & 64.6/53 & 100.1/53 \\
BR($\tpppc$) & 9.11\% & 9.13\%  & 9.21\% \\
 \hline
  \end{tabular}
  \caption[]{Results of three-parameter fits to ALEPH data 
  for fixed values of $\alpha$. $M_A$ and $\Gamma_{a_1}(M_A^2)$ are 
  given in units of GeV.}
\label{tab2}
\end{center}
\end{table}

\begin{figure}[t]
\vspace*{0.6cm}
\centerline{
   \includegraphics[width=11cm,angle=-90]{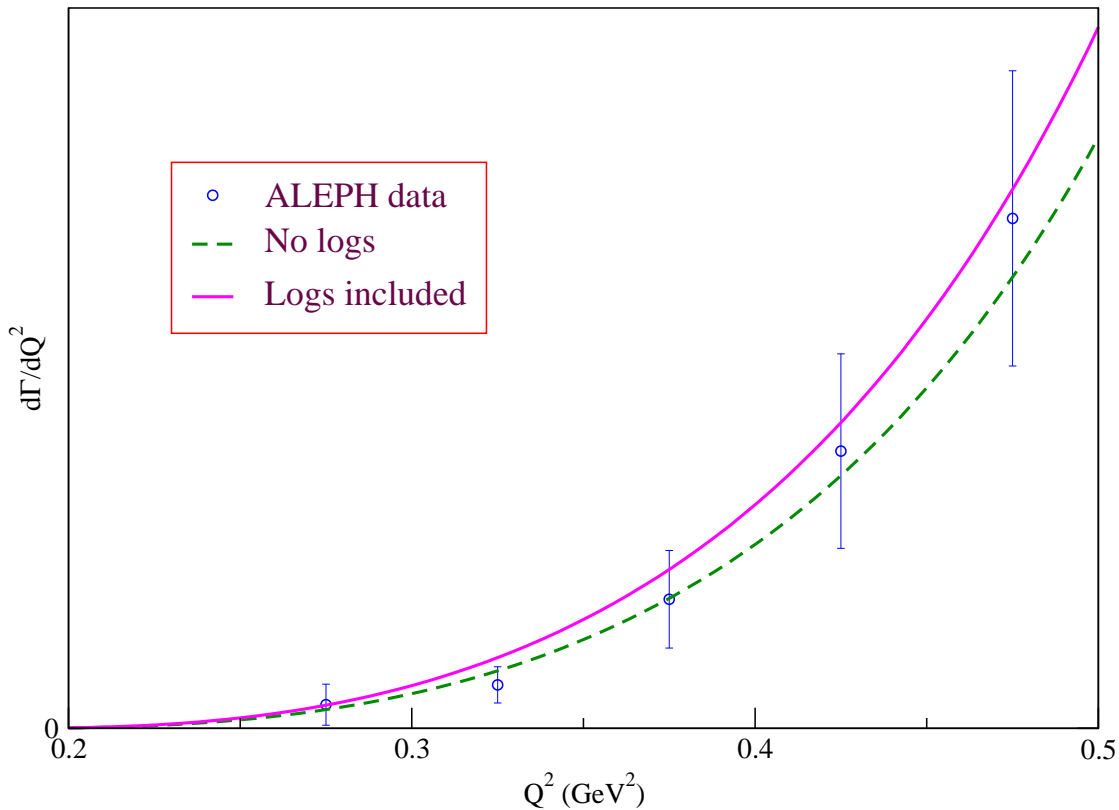}
   } \caption{Theoretical curves for $\tpppc$ spectrum at $Q^2 \ll M_V^2$,
    with and without
the inclusion of chiral logs, vs.\ ALEPH data in the low $Q^2$ region
\protect{\cite{ALE}}.}
\label{fig:cuarta}
\end{figure}

\begin{figure}[t]
\centerline{
   \includegraphics[width=13truecm,angle=-90]{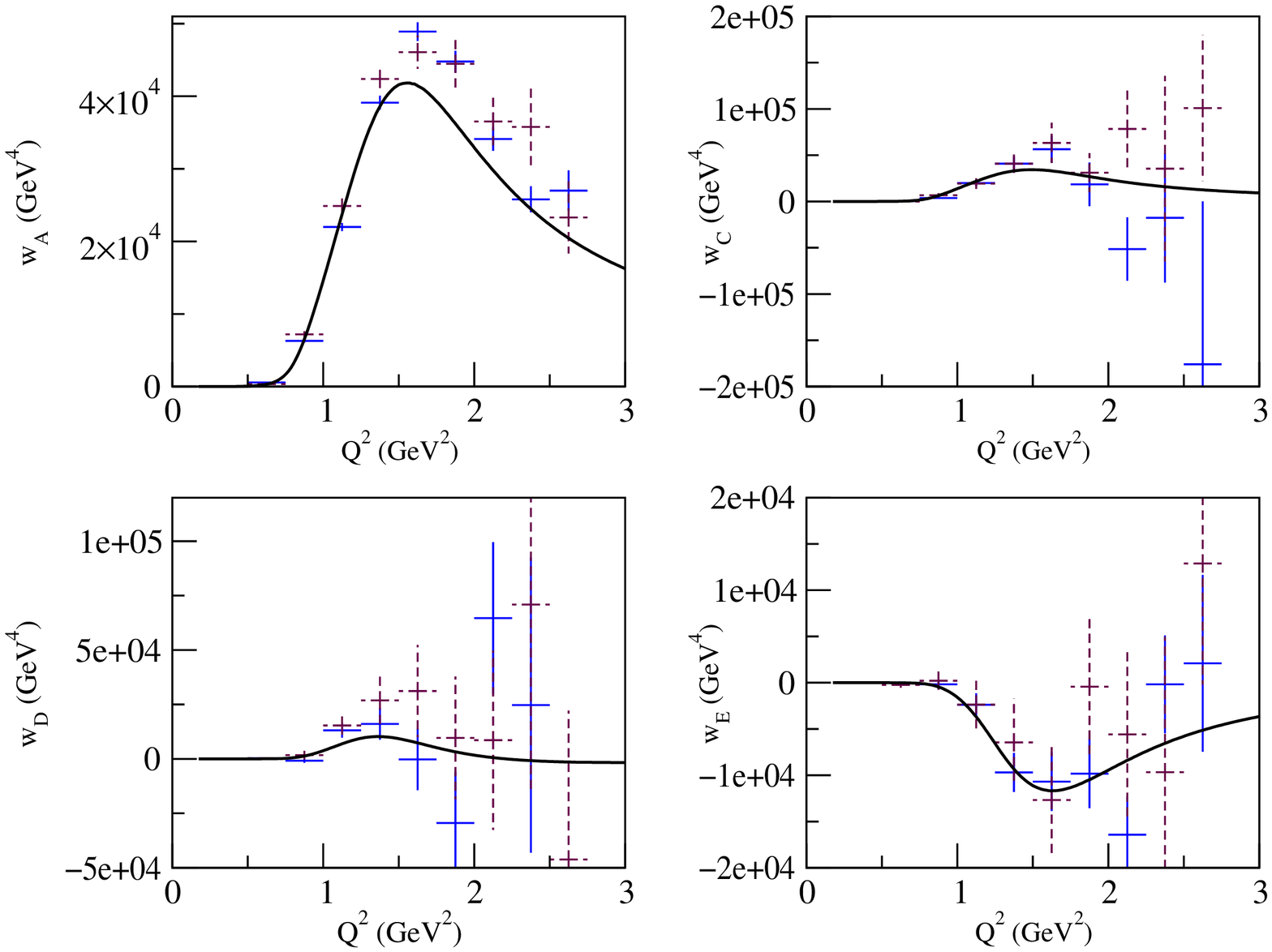}
   }
   \vspace*{0.2cm}
   \caption{Theoretical values for the $w_A$, $w_C$, $w_D$ and 
   $w_E$ integrated structure functions 
   in comparison with the experimental data from CLEO-II (solid) and 
   OPAL (dashed) \protect{\cite{WXexp}}. All of them have been normalized
   to their respective branching ratios. }
\label{fig:quinta}
\end{figure}

To conclude this analysis, we take a closer look to the low 
$Q^2$ region
of the spectrum. In fact, in our approach we have assumed that ${\cal
O}(p^4)$ corrections arising from chiral logs are small, hence the
dominant contributions to hadronic amplitudes arise from resonance
exchange. Close to threshold (i.e.\ for $\sqrt{Q^2}$ well below $M_V$) 
one is able to explicitly calculate the contributions of
${\cal O}(p^4)$ chiral logs, therefore our theoretical prediction for the
spectrum in this region can be improved, and the impact of chiral logs can
be numerically evaluated. In order to add these contributions to the
hadronic amplitude, we have made use of the calculations performed in
Ref.\ \cite{CFU:96} within standard $\chi$PT. The corresponding results are
sketched in Appendix A. The effect on the $\tpppc$ spectral function is
shown in Fig.\ \ref{fig:cuarta}, where we display both the 
experimental values from ALEPH
and the curves for the spectrum with and without the inclusion of ${\cal
O}(p^4)$ chiral logs. It can be seen that the corrections produce a slight
enhancement of the spectral function. However, given the size of the
experimental errors, the quality of the agreement with ALEPH data is
unchanged.
\par
When enforcing the asymptotic conditions of QCD in Section~4, we pointed
out that the relation in Eq.~(\ref{eq:dubio}) is not necessarily true when 
the vertices in ${\cal L}_2^{\rm VAP}$ are considered, because the latter
contribute to the axial form factor. If we do not take into account
that constraint we can perform a 5 parameter fit to the 
ALEPH data that gives~:
\begin{eqnarray}
\label{fito}
\lambda_0 & = & 11.0 \pm 1.7 \; \; , \nonumber \\
\alpha & = & 2.55 \pm 0.15 \; \; , \nonumber \\
M_{A} & = & (1.19 \pm 0.02) \rm{\ GeV} \; \; , \nonumber \\
\Gamma_{a_1}(M_A^2) & = & (0.47 \pm 0.02) \rm{\ GeV} \; \;, \nonumber \\
\Frac{G_V}{F_V} & = & 0.49 \pm 0.03  \; \; \; ,
\end{eqnarray}
and $\chi^2/\rm{d.o.f.} = 63.7/51$. We see that the ratio $G_V/F_V$
is consistent with the value given by Eq.~(\ref{eq:dubio}) within 
the error. Hence we confirm that the
experimental data favour the $F_V = 2 G_V$ constraint.
\par
Finally a word of caution has to be said about the value of $\lambda_0$.
The only dependence of the 
amplitude on this parameter is given by the function $F(Q^2,s)$ in
Eq.~(\ref{eq:lambda0}) where, as can be seen, $\lambda_0$ is multiplied
by the squared mass of the pion. Hence the sensibility of the observables
in $\tppp$ on $\lambda_0$ is very much covered up. Tau decays into kaons
would provide a better testing field to find out on $\lambda_0$
and check the results against the values from our fits. We
consider that our result on this parameter has to be taken with care. 

\subsection{Description of structure functions}

Structure functions provide a full description of the hadronic tensor
$T_{\mu} T_{\nu}^*$ in the hadron rest frame. There are 16 real
valued structure functions in $\tau \rightarrow P_1 P_2 P_3 \,\nu_{\tau}$
decays ($P_i$ is short for a pseudoscalar meson), most of which can be
determined by studying angular correlations of the hadronic system. 
Four of them carry information on the $J^P=1^+$
transitions only~: $w_A$, $w_C$, $w_D$ and $w_E$ (we refer the reader to 
Ref.\ \cite{KM:92} for their precise definitions). 
Indeed, for the $\tppp$ processes, 
other structure functions either vanish identically, or involve the 
pseudoscalar form factor $F_P$, which appears to be strongly suppressed above 
the very low--energy region due to its proportionality to the squared
pion mass. 
\par
Both CLEO-II and OPAL have measured the four structure
functions quoted above for the $\tpppn$ process, while concluding
that other functions are consistent with zero within errors. 
Hence we can proceed to compare those experimental results with 
the description that provides our theoretical approach. 
In our expressions for the structure functions we input the values
of the parameters obtained from the fit in Eq.~(\ref{fit}), getting 
the theoretical curves shown in Fig.\ \ref{fig:quinta}. The latter
are compared with the
experimental data quoted by CLEO and OPAL \cite{WXexp}. For
$w_C$, $w_D$ and $w_E$, it can be seen that we get a good agreement
in the low $Q^2$ region, while for increasing energy the experimental
errors become too large to state any conclusion (moreover, there seems
to be a slight disagreement between both experiments at some points).
\par
On the other hand, in the case of the integrated structure function $w_A$,
the quoted experimental errors are smaller, and the theoretical curve seems
to lie somewhat below the data for $Q^2 \gsim 1.5 \, \mbox{GeV}^2$. 
However, it happens that $w_A$
contains essentially the same information about the hadronic amplitude as
the spectral function $d\Gamma/dQ^2$. This becomes clear by looking at
Eq.\ (\ref{dgdq}) if the scalar structure function $W_{SA}$ is put to zero
(remember that it should be suppressed by a factor 
${\cal O}(m_\pi^2/Q^2)$). Taking into account that $w_A$ is given by
\begin{equation}
w_{A}(Q^2) = \int ds\, dt\; W_A(Q^2,s,t)\, \, \;,
\end{equation}
where $W_A$ is the structure function previously introduced in Eq.\
(\ref{wa}),  one simply has
\begin{equation}
\label{specwa}
\Frac{d\Gamma}{dQ^2} \,  = \,  \frac{G_F^2\,|V_{ud}|^2}
{384\, (2\pi)^5\, M_\tau}\;
\left( \, \Frac{M_\tau^2}{Q^2}-1 \, \right)^2 
\left(1 \, + \, 2 \,\frac{Q^2}{M_\tau^2}\right) \; \; w_A(Q^2)  \; \; .
\end{equation}
\begin{figure}[t]
\begin{center}
   \includegraphics[width=12truecm,angle=-90]{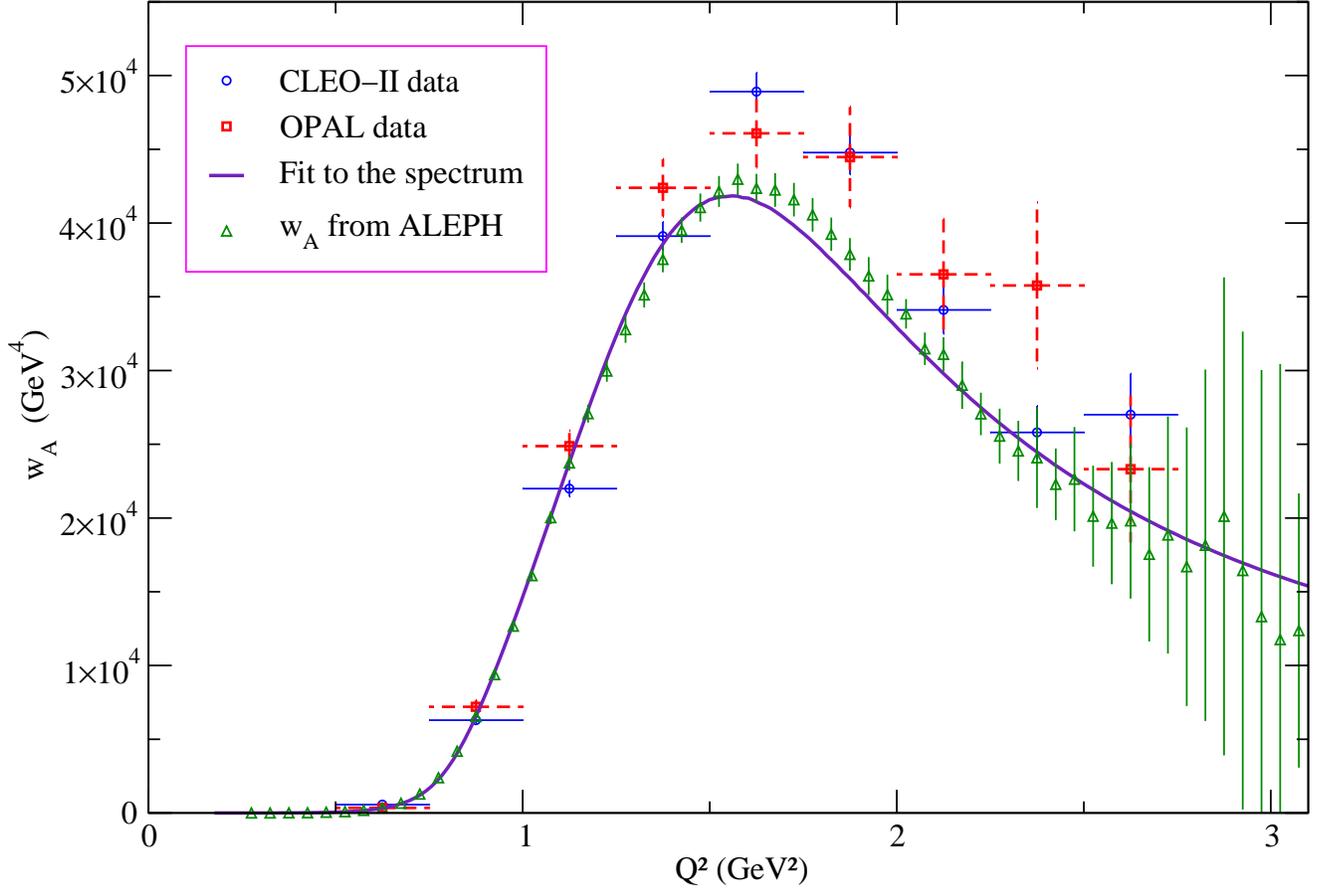}
\end{center}
\vspace*{-0.4cm}
 \caption{Comparison between the experimental data for $w_A$, from
   $\tpppn$, quoted by
CLEO-II and OPAL \protect{\cite{WXexp}} and the values arising from 
ALEPH measurements of $\tpppc$ spectral functions \protect{\cite{ALE}}. 
The solid line is obtained from our four-parameter fit of the spectrum 
(see text).}
\label{fig:sexta}
\end{figure}

\begin{figure}[!tbp]
\begin{center}
   \includegraphics[scale=0.8]{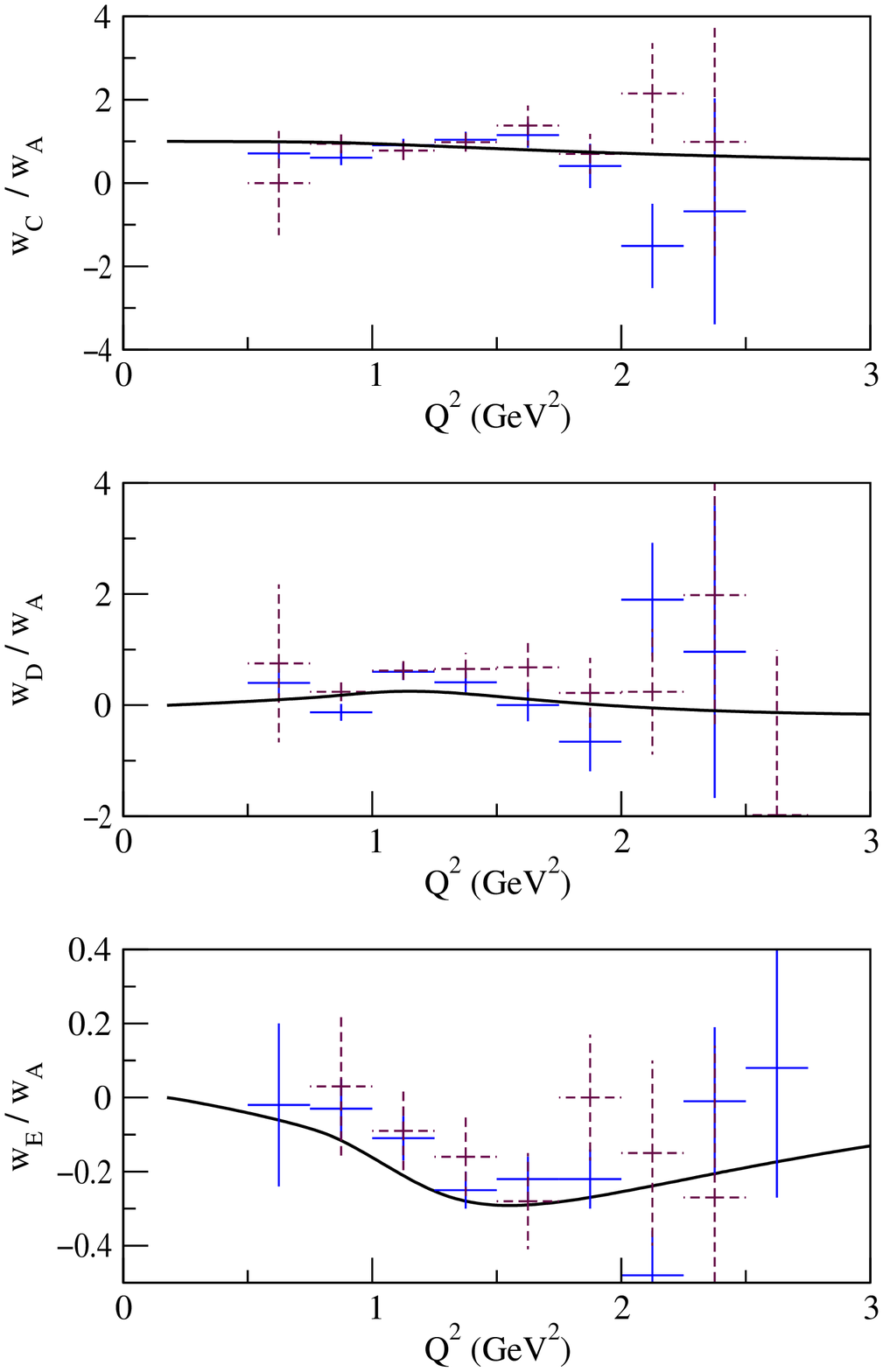}
\end{center}
\caption{Experimental data for $w_C$, $w_D$ and $w_E$ integrated
structure functions, normalized to $w_A$. Solid (dashed) bars correspond
to the values quoted by CLEO-II (OPAL) \protect{\cite{WXexp}}.}
\label{fig:septima}
\end{figure}

In this way  one can compare the
measurements of $w_A$ quoted by CLEO-II and OPAL with the data obtained by
ALEPH for the spectral function, conveniently translated into $w_A$. This
is represented in Fig.\ \ref{fig:sexta}, where it can be seen that some of the
data from the different experiments do not agree with each
other within errors. Notice that, due to phase space suppression, the 
factor of proportionality between $w_A(Q^2)$ and $d \Gamma / d Q^2$ in 
Eq.~(\ref{specwa}) goes to zero for $Q^2 \rightarrow M_{\tau}^2$, therefore
the error bars in the ALEPH points become enhanced toward the end
of the spectrum. Notwithstanding, up to $Q^2\simeq 2.5$ GeV$^2$,
it is seen that ALEPH errors are still smaller than those corresponding to
the values quoted by CLEO-II and OPAL. On this basis, we have chosen here to
take the data obtained by ALEPH to fit the unknown theoretical parameters
in the hadronic amplitude. Finally, notice that a non vanishing
contribution of $W_{SA}$ (which is a positive quantity) cannot help to
solve the experimental discrepancies, as it would go in the wrong direction.
\par
In the analysis of data carried out by the CLEO Collaboration \cite{cleofa}
onto their $\tpppn$ results it was concluded that the data was showing large
contributions from intermediate states involving the isoscalar
mesons $f_0$(600), $f_0$(1370) and $f_2$(1270). Their analysis was done
in a modelization of the axial--vector form factors that included 
Breit--Wigner functions in a K\"uhn and Santamaria inspired model. Our
results in the Effective Theory framework show that, within the present 
experimental errors, there is no evidence of relevant contributions in 
$\tppp$ decays beyond those of the $\rho$(770) and $a_1$(1260) resonances. 
\par
We also find interesting to study the integrated structure functions
$w_C$, $w_D$ and $w_E$, normalized to $w_A$. This quantities should be
less sensitive to the parameterization of the $a_1$(1260) width, and do not
depend on the global normalization used by each experiment (which could be
the origin of the partial discrepancies in the values for $w_A$). The
corresponding comparison between the theoretical and experimental 
results are plotted in Fig.\ \ref{fig:septima}, where once again it 
is seen that our 
predictions are in good agreement with the experimental data within their
present errors.

\section{Conclusions}

The hadronization of QCD currents yields relevant information on 
non--perturbative features of low--energy Quantum Chromodynamics.
At $E \ll M_V$, $\chi$PT has provided model--independent knowledge
on the corresponding form factors, but when resonance degrees of freedom
become active, at $E \sim M_V$, they have to be properly included 
into the Effective Theory and $\chi$PT is no longer the appropriate
scheme to work with. While thorough parameterizations of the hadron 
matrix amplitudes involving resonances have been used in the past, 
most of them implement Breit--Wigner functions in order to provide
the dynamical description of the resonances and, as far as we know,
there is no deductive connexion between those functions and QCD.
Hadronic $\tau$ decays take place in the bulk of the energy resonance
region and, moreover, several experiments in the last years have
payed attention to them, supplying a good quality and quantity of 
data. Therefore they constitute an interesting subject of research
to which we have turned over.  
\par
In this article we have proposed an Effective Theory approach in order
to evaluate the dominant form factors in $\tppp$ decays, and our results
have been compared with the experimental figures. 
To proceed to the construction of the relevant Effective Theory we
have relied in the R$\chi$T proposed in Ref.~\cite{EGPR:89}, which has been
improved here by implementing the necessary vector--axial--vector--Goldstone
interactions ${\cal L}_2^{\rm VAP}$ in Eqs.~(\ref{eq:lag21},
\ref{lag2}). The basic principles underlying this procedure are the
all--important chiral symmetry of massless QCD, that drives the 
interaction of Goldstone bosons at low--energy, and the unitary 
flavour symmetry of the octets of resonances. 
Once specified our interaction Lagrangian we have proceeded to evaluate 
the axial--vector current form factors in $\tppp$ processes. 
It is clear though that, up to this point, the Lagrangian 
${\cal L}_{\rm R\chi T}$ is not a proper Effective Theory of 
QCD for arbitrary values of the coupling constants. Hence we have
implemented the consequences of the asymptotic behaviour of 
vector and axial--vector spectral functions of the underlying QCD
into the corresponding form factors. This procedure has fixed up 
several constraints onto the couplings, yielding an Effective
Theory more germane to QCD.  
\par
After applying this method we are left with four unknown parameters
in our results for the axial--vector form factors. We have used
the ALEPH data on $\tpppc$ to perform a fit to its branching ratio
and spectral function, rendering values for the mass and on--shell
width of the $a_1$(1260) resonance. Once the parameters in the
theoretical expressions of the form factors have been determined, we 
have evaluated
the integrated structure functions $w_A$, $w_C$, $w_D$ and $w_E$
measured by the CLEO-II and OPAL experiments \cite{WXexp}. We have found
a good description of data, though there appear to be some inconsistencies
between the results provided by the different experiments.
\par
In summary we have provided an Effective Theory based evaluation of 
the axial--vector form factors in $\tppp$ decays that gives an 
appropriate account of the main features of the experimental data. 
The procedure does not rely in any modelization of the form factors,
as it has been done in the past, but in a field theory construction 
that embodies the relevant features of QCD in the resonance energy
region, showing that this is a compelling framework to work with.

\section*{Acknowledgements}

D.G.D.\ acknowledges financial support from Fundaci\'on Antorchas and
CONICET (Argentina). J.~Portol\'es is supported by a \lq \lq Ram\'on y Cajal"
contract with CSIC funded by MCYT.
This work has been supported in part by TMR EURIDICE, EC Contract No. 
HPRN-CT-2002-00311, by MCYT (Spain) under grant FPA2001-3031, by
Generalitat Valenciana under grant GRUPOS03/013, by
the Agencia Espa\~nola de Cooperaci\'on Internacional (AECI) and
by ERDF funds from the European Commission.

\appendix
\newcounter{erasmo}
\renewcommand{\thesection}{\Alph{erasmo}}
\renewcommand{\theequation}{\Alph{erasmo}.\arabic{equation}}
\renewcommand{\thetable}{\Alph{erasmo}}
\setcounter{erasmo}{1}
\setcounter{equation}{0}
\setcounter{table}{0}

\section{The ${\cal O}(p^4)$ $\chi$PT result for $\tpppc$}

In the very low $Q^2$ region (typically $E \ll M_V$), our results for 
the hadronic amplitudes in $\tppp$ can be compared with those obtained 
in standard $\chi$PT. Here we will
rely on the analysis in Ref.\ \cite{CFU:96}, where these calculations have
been performed in detail up to ${\cal O}(p^4)$.
\par
The ${\cal O}(p^4)$ hadronic matrix
element $T_\mu$ involves renormalized coupling constants $\ell_i^r(\mu)$,
which have to be determined experimentally at a given renormalization
scale. It is seen that these couplings can be related to finite and
scale-independent quantities $\bar \ell_i$ \cite{GL:85} according to~:
\begin{equation}
\bar \ell_i = \left(\frac{\gamma_i}{32\pi^2}\right)^{-1} \ell_i^r(\mu)
 - \ln \frac{m_\pi^2}{\mu^2}\;,
\end{equation}
where the coefficients $\gamma_i$ arise from the corresponding
renormalization group equations. In the resonance chiral effective theory,
the constants $\bar \ell_i$ are assumed to be saturated by resonance
exchange, which at low energies induces the local $\chi$PT Lagrangian of
${\cal O}(p^4)$ for the light pseudoscalar mesons \cite{EGPR:89}. In this
way, after considering the relations (\ref{fvgv}), imposed by QCD, one gets
\begin{eqnarray}
\bar \ell_1 & = & -48\pi^2 \frac{F^2}{M_V^2}
-\ln\frac{m_\pi^2}{M_V^2} \; \; , \nonumber \\
\bar \ell_2 & = & 24\pi^2 \frac{F^2}{M_V^2}
-\ln\frac{m_\pi^2}{M_V^2} \; \; , \nonumber \\
\bar \ell_4 & = & -\ln\frac{m_\pi^2}{M_V^2} \; \; , \nonumber \\
\bar \ell_6 & = & 96\pi^2 \frac{F^2}{M_V^2}
-\ln\frac{m_\pi^2}{M_V^2}\; \;.
\end{eqnarray}
Now, from the expressions quoted in Ref.\ \cite{CFU:96}, the $J^P=1^+$ piece
of the hadronic amplitude for the decay $\tpppc$ is found to be
\begin{eqnarray}
\left. T_{+\mu}^{\chi PT}\right|_{1^+}\, & = &
\, - \,\frac{2\sqrt{2}}{3F}\, V_{1\mu}\,
\Bigg\{ 1 + \frac{3s}{2M_V^2} + \frac{1}{32\pi^2 F^2}
\bigg[-\frac{2}{3}\, s + \, 3\, t - u -3\, m_\pi^2
- (s + m_\pi^2)\, F(s/m_\pi^2) \nonumber \\
& & \; \; \; \; \; \; \; \; \; \; \; \; \; \; \; \; \; \; 
\; \; \; \; \; \; \; \; \; \; \; \; \; \; \; \; \; \; 
\; \; \; \; \; \; \; \; \; \; \; \; \; \;  
 + (3\,t-2\,m_\pi^2)\, F(t/m_\pi^2)
- (u-2\,m_\pi^2)\, F(u/m_\pi^2) \nonumber \\
& & \; \; \; \; \; \; \; \; \; \; \; \; \; \; \; \; \; \; 
\; \; \; \; \; \; \; \; \; \; \; \; \; \; \; \; \; \; 
\; \; \; \; \; \; \; \; \; \; \; \; \; \;    
+ (s-3t+u-m_\pi^2)\,\ln
\frac{m_\pi^2}{M_V^2}\,\bigg]\Bigg\} \nonumber \\
& & \, - \,\frac{2\sqrt{2}}{3F}\, V_{2\mu}\,
\Bigg\{\;\;s\longleftrightarrow t\;\; \Bigg\} + {\cal O}(p^6) \;\;,
\end{eqnarray}
where
\begin{equation}
F(x) = \sigma \, \ln \frac{1-\sigma}{1+\sigma}\;\;,
\hspace{1cm}
\sigma(x) = \sqrt{1-4/x}\;\;,
\end{equation}
and the kinematical invariants $s$ and $t$ are defined as in 
Section 3.
The first two terms in the curly brackets are those explicitly quoted in
Eq.\ (\ref{chpt}).

\end{document}